%% file: conference_101719.tex
\documentclass[conference]{IEEEtran}
\IEEEoverridecommandlockouts
\usepackage{amsmath,amssymb,amsfonts}
\usepackage{algorithm, algorithmic}
\usepackage{graphicx}
\usepackage{textcomp}
\usepackage{xcolor}
\usepackage{subcaption}
\usepackage{comment}
\usepackage{multirow}
\usepackage{adjustbox, trimclip}
\usepackage{tikz, pgfplots}
\usepackage{booktabs}
\usetikzlibrary{arrows.meta, shapes.multipart, calc, positioning, shapes.misc, patterns.meta, patterns}
\pgfplotsset{compat=1.18}

\def\BibTeX{{\rm B\kern-.05em{\sc i\kern-.025em b}\kern-.08em
    T\kern-.1667em\lower.7ex\hbox{E}\kern-.125emX}}

\DeclareCaptionFormat{ieee_table}
{%
    \textsc{#1#2\\#3}
}

\DeclareCaptionLabelSeparator{ieee_table}
{} 

\usepackage[
    backend=biber,
    style=ieee,
    maxbibnames=999,
    maxcitenames=999,
    defernumbers=true,
    doi=false,
    url=false,
]{biblatex}

\DefineBibliographyStrings{english}{
    june = June,
    july = July,      
}

\AtBeginBibliography{\footnotesize}

\begin{document}

\title{High-throughput Low-latency Hardware Implementation of BCH Decoders\\
\thanks{The work of J. Lagendijk, Y. C. G{\"u}ltekin, A. Balatsoukas-Stimming and A. Alvarado is part of the project BIT-FREE with file number 20348 of the research programme Open Technology Programme which is (partly) financed by the Dutch Research Council (NWO).}
}
\author{Jasper Lagendijk\textsuperscript{\textdagger,*}, Wenqing Song\textsuperscript{\textdaggerdbl}, Yunus Can G\"ultekin\textsuperscript{\textdagger},\\ Andreas Burg\textsuperscript{\textdaggerdbl}, Alex Alvarado\textsuperscript{\textdagger}, Alexios Balatsoukas-Stimming\textsuperscript{\textdagger} \\
\textit{\textsuperscript{\textdagger}Department of Electrical Engineering, Eindhoven University of Technology} \\
\textit{\textsuperscript{\textdaggerdbl}Department of Electrical Engineering, École Polytechnique Fédérale de Lausanne} \\
\textsuperscript{*}j.lagendijk@tue.nl
}
\maketitle

\begin{abstract}
Two well-known decoding algorithms for BCH codes are conventional decoding, based on the Berlekamp-Massey algorithm in combination with Chien search, and direct decoding, which uses direct solutions to find the error locator polynomial and its roots. 
We introduce hardware architectures for conventional and direct decoding of extended BCH codes. 
Both architectures support implementation for any blocklength.  
Our conventional decoder supports any error-correction capability, whereas direct decoding is supported up to error correcting capability $t=4$.  
To the best of our knowledge, our work is the first to implement a direct BCH decoder with an error-correction capability 4.
We synthesize for the Xilinx Ultrascale+ XCZU48DR field-programmable gate-array and 16~nm FinFET for blocklengths up to $1024$ bits and $t=4$. 
We show that the direct decoder outperforms the conventional decoder in area efficiency for $t=2$, $t=3$, and for $t=4$ for blocklengths longer than $256$.
Post-synthesis results for 16~nm FinFET show codeword per clock-cycle throughput at 1 GHz, achieving 239 Gb/s for the $(256, 239)$ eBCH code and 223 Gb/s for $(256, 223)$ eBCH code at $2$~ns and $8$~ns latency, respectively.
\end{abstract}


\input{Sections/introduction}
\input{Sections/bch_codes}

\input{Sections/hardware_implementation}
\input{Sections/performance_analysis}
\input{Sections/conclusions}
\newpage
\printbibliography

\end{document}

%% file: Sections/introduction.tex
\section{Introduction}
Modern data-center interconnects are required to achieve extremely high throughput at low latency and power consumption. Well-designed efficient 
forward error correcting~(FEC) codes and decoders are thus of great importance. Generalized product codes (GPCs) with Bose-Chaudhuri-Hocquenghem~(BCH) and extended BCH~(eBCH) component codes are used widely in fiber optic communications and are strong candidates for next generation optical interconnects~\cite{Jian2013GLOBECOM, Xie2021ICTA}. 
These spatially coupled codes combine the low complexity of decoding the component (e)BCH codes and the high reliability achieved by the iterative decoding on the spatially coupled structure~\cite{Fougstedt2019JLT}. 

BCH codes are algebraic linear block codes defined for two parameters, a field power $m$ and an error correcting capability $t$~\cite{Bose1960IaControl,Hocquenghem1959Chiffres}. 
These two parameters determine the blocklength $n=2^m-1$ and information length $k\geq n-mt$. 
eBCH codes are BCH codes extended by an additional parity bit, guaranteeing $t+1$ error detection. 
Decoding of binary BCH codes consists of three steps: calculating the syndrome components, computing the error locator polynomial (ELP), and finding the roots of the ELP. 
Conventional decoding of BCH codes is based on computing the ELP using the Berlekamp-Massey (BM) algorithm in combination with Chien search. 
An alternative is computing the roots directly, which is possible for polynomials up to $t=4$~\cite{Hamming1950BellSTJ,Polkinghorn1966TIT,Yan1998EL}. 
We refer to this decoding strategy as direct decoding. 

Many hardware implementations for BCH decoders have been proposed~\cite{ZhangISCAS2018, Choi2019VLSI, Fougstedt2019JLT, Ou2025CICC, Wei2024SSCL, Park2018ACCESS}. 
These implementations are designed for a specific use case, e.g., they are only designed for specific blocklengths and/or error correcting capabilities~\cite{Choi2019VLSI, Fougstedt2019JLT, ZhangISCAS2018}, they are designed using serial implementations of polynomial multiplication and division~\cite{Ou2025CICC}, or they are designed to handle many different Galois fields at runtime, which comes with a significant increase in complexity~\cite{Wei2024SSCL,Park2018ACCESS}. 
To our knowledge, no prior work has presented a highly parallel, low-latency hardware architecture for a conventional eBCH decoder that supports arbitrary $n$ and $t$ at compile time, while delivering a throughput of one codeword per clock cycle. 
Architectures for the direct algorithm exist only up to $t=3$, no implementations for $t=4$ exist. 

\emph{Contributions:} In this work, we implement both a conventional eBCH decoder, using BM and Chien search, for any $t$ and a direct decoder for $t\leq4$ using a high-throughput, low-latency architecture. Our decoders are fully pipelined, achieving one codeword per clock cycle throughput. 
To the best of our knowledge, our work is the first architecture of a direct decoder with $t=4$. 
For our implementation, the parameters $n$ and $t$ are chosen at compile time, they are constant at runtime. 
We then analyze these architectures for a range of blocklengths $n$.
For the FPGA target, we report post-implementation (place-and-route) results on a Xilinx Ultrascale+ XCZU48DR FPGA~\cite{AMD:ZynqUltraScale+}. For the ASIC target, we report post-synthesis results using a 16~nm FinFET standard-cell library. We analyze the post-synthesis area, throughput, and latency for $n\leq1024$ and $t\leq4$.

The remainder of this paper is organized as follows. In Section~\ref{BM_chien} we introduce the Berlekamp-Massey algorithm and Chien search. Similarly, in Section~\ref{direct_algorithm}, we explain the direct decoding algorithms. In Section~\ref{hardware_architecture} we describe our proposed hardware architectures for these algorithms. The implementation results for our hardware architectures are then analyzed in Section~\ref{results}. Conclusions are provided in Section~\ref{conclusions}.

%% file: Sections/bch_codes.tex
\section{Conventional decoding of BCH codes}
\label{BM_chien}
The conventional BCH decoder consists of three stages: (1) computing the syndrome components, (2) computing the ELP through BM, and (3) finding the roots of the ELP through Chien search. 
This method is slower and more computationally costly compared to direct implementation~\cite{Sukmadji2024TC}. However, this method can be used for any $t$, while the direct method is limited to $t\leq4$. 
In this paper we focus exclusively on binary BCH codes. 

Given a received sequence $\mathbf{r}$ of $n$ bits, the first step is to compute the syndrome components, $S_1, \cdots,S_{2t}$, to pass to the BM algorithm. These syndrome components are entries in the Galois field. Traditionally, these syndromes are computed using minimal polynomials, polynomial division and serial linear feedback shift registers~\cite[Sec. 6.8]{ErrorControlCoding:2004}. For binary BCH codes, direct matrix multiplication is a viable alternative. The syndrome components can then be calculated according to
\begin{equation}
    \mathbf{S} = \mathbf{r}\mathbf{H}^T, \qquad \mathbf{S}=(S_1, S_2,\dots,S_{2t}),
\end{equation}
where $\mathbf{H}$ is the $2t\times n$ parity check matrix.
These syndrome components are then passed to the BM algorithm, to compute the ELP.  

The BM algorithm is an iterative algorithm for finding the ELP of a $t$ error-correcting BCH code in $2t$ iterations~\cite{Berlekamp1965TIT,Massey1969TIT}. 
In the case of binary BCH codes, BM can be simplified to $t$ iterations~\cite[Ch. 6.4]{ErrorControlCoding:2004}, exploiting the property that $S_i^2 = S_{2i}$ for binary BCH codes. 
The computation of the ELP $\Lambda(X)$ using simplified BM, based on $2t$ input syndrome components, consists of three parts, as shown in Algorithm~\ref{alg:berlekamp_massey}. 
First (lines \ref{alg_line:init_start}-\ref{alg_line:init_end}), the decoder is initialized. Then for $0\leq\mu < t$ the polynomial $\Lambda^{(\mu+1)}(X)$ is computed iteratively.  
After $\mu$ iterations, a potential ELP $\Lambda^{(t)}(X)$ is found. If $\text{deg } \Lambda^{(t)}(X) \leq t$ the polynomial is accepted and Chien search can start. If not, the decoder declares a failure.

The BM algorithm requires the values from two iterations, the current iteration $\mu$ and a secondary iteration $\rho$, with $\rho<\mu$, to compute the next iteration $\mu+1$. 
During initialization, the values for two iterations are set, $\mu=-1/2$ and $\mu=0$;
the polynomials $\Lambda^{\left(-\frac{1}{2}\right)}(X)$ and $\Lambda^{(0)}(X)$ are set to $1$ and the discrepancies, which can be described as the prediction error for iteration $\mu$, are set to $d^{(-\frac{1}{2})}=1$ and $d^{(0)}=S_1$. 
For iteration $\mu=0$, $\rho=-1/2$ is used. The use of a half-iteration $\rho=-1/2$ is an effect of the simplification from $2t$ to $t$ iterations, the stride $X^{2(\mu-\rho)}$ in line \ref{alg_line:bm_mu} is jointly motivated by this convention. 
After initialization, the iterative algorithm starts (lines \ref{alg_line:bm_start}-\ref{alg_line:bm_end}). 
For $\mu=0$ to $\mu=t-1$, first the ELP for $\mu+1$ is computed (line \ref{alg_line:bm_mu})
and the discrepancy is computed (line \ref{alg_line:bm_discrepancy}). 
Before continuing from $\mu$ to $\mu+1$, a check is made whether iteration $\mu$ is a valid candidate to be selected as the new previous iteration $\rho$. There are two requirements for this: $d^{(\mu)}\neq0$ and $2\rho - l^{(\rho)} < 2\mu-l^{(\mu)}$, where $l^{(\mu)}=\text{deg }\Lambda^{(\mu)}(X)$. If both of these conditions are satisfied, then $\rho$ is set to $\mu$. Afterwards, the next iteration can start. 
After $t$ iterations, if a valid polynomial is found, Chien search can be applied to the polynomial. 
 
\begin{algorithm}[t]
 \caption{Simplified Berlekamp-Massey}
 \label{alg:berlekamp_massey}
 \begin{algorithmic}[1]
    \STATE \textbf{Input} $S_i  \text{ for } 1\leq i \leq 2t$
    \STATE \textbf{Output} $\Lambda(X)$ or decoding failure
    \STATE $\mu \leftarrow 0$ \label{alg_line:init_start}
    \STATE $\rho \leftarrow -\frac{1}{2}$
    \STATE $d^{(-\frac{1}{2})}=1$
    \STATE $d^{(0)} = S_1$
    \STATE $l^{(-\frac{1}{2})} \leftarrow 0$
    \STATE $l^{(0)} \leftarrow 0$
    \STATE $\Lambda^{(-\frac{1}{2})}(X) \leftarrow 1$
    \STATE $\Lambda^{(0)}(X) \leftarrow 1$ \label{alg_line:init_end}
    \WHILE{$\mu < t$} \label{alg_line:bm_start}
        \STATE $\Lambda^{(\mu+1)}(X) \leftarrow \Lambda^{(\mu)}(X) + \frac{d^{(\mu)}}{d^{(\rho)}} X^{2(\mu-\rho)}\Lambda^{(\rho)}(X) $ \label{alg_line:bm_mu}
        \STATE $d^{(\mu+1)}\leftarrow \sum_{i=0}^{\mu}S_{{2}\mu+3-i}\Lambda_i^{(\mu+1)},\text{ } (S_j=0 \text { for } j>2t)$\label{alg_line:bm_discrepancy}
        \STATE $l^{(\mu+1)} \leftarrow \text{ deg } \Lambda^{(\mu+1)}(X)$
        \IF{$d^{(\mu)}\neq0$ and $2\rho - l^{(\rho)}<2\mu-l^{(\mu)}$}
            \STATE $\rho\leftarrow \mu$
        \ENDIF
        \STATE $\mu\leftarrow \mu+1$
    \ENDWHILE \label{alg_line:bm_end}
    \IF{$\text{deg } \Lambda^{(t)}(X) > t$}
        \STATE Decoding failure 
    \ELSE 
        \STATE $\Lambda(X)=\Lambda^{(t)}(X)$
    \ENDIF
 \end{algorithmic}
\end{algorithm}

Chien search is a brute force method of finding the roots of a polynomial in a Galois field.
Roots are found by checking if $\Lambda(X)=0$ for all $X=\alpha^i$, with $0 \leq i < n$. 
For our ELP $\Lambda(X)$, if $\Lambda(\alpha^i)=0$, a root is found and the error locations can be computed. 
For a given root $\alpha^i$, the corresponding error location $e_i$ can be found using $ e_i = (n-i) \text{ mod } n$.


\section{Direct Root Finding}
\label{direct_algorithm}
General direct solutions for finding the root of polynomials in GF($2^m$) exist for polynomials up to degree $4$. 
For larger degree polynomials, no direct solutions exist and brute force methods, like Chien search, have to be used. 
For binary BCH codes, direct solutions have been shown for $t=1$ in~\cite{Hamming1950BellSTJ}, $t=2,3$ in~\cite{Polkinghorn1966TIT} and for $t=4$ in~\cite{Yan1998EL}. 
For all four versions, the decoding process is divided into three parts: (1) the needed syndrome components are computed (only the first $t$ odd syndrome components are needed for binary BCH codes), 
(2) these syndrome components are then used to construct the ELP, and (3) root finding techniques are used from which the error locations can be computed.

Any degree-$t$ polynomial is given by
\begin{equation}
    \Lambda(X) = \Lambda_0 + \Lambda_1X + \cdots + \Lambda_tX^t.
\end{equation}
The main method for finding the roots of the polynomial $\Lambda(X)$ is based on transforming the polynomial into either the form
\begin{equation}
    A(X)=X^2 + X + k,
    \label{eq:second_order_solvable}
\end{equation}
or into the form
\begin{equation}
    B(X)=X^3+X+k,
    \label{eq:third_order_solvable}
\end{equation}
or into the form
\begin{equation}
    C(X) =X^3+k.
    \label{eq:third_order_solvable_no_linear}
\end{equation}
For a given $k$, if roots exist, they can be found using a lookup table (LUT) with  $2^m-1$ entries with two outputs for \eqref{eq:second_order_solvable}, or three outputs for \eqref{eq:third_order_solvable} and \eqref{eq:third_order_solvable_no_linear} each. We denote the application of these root finding LUTs as $\{\}_A$, $\{\}_B$, and $\{\}_C$, respectively. It should be noted that \eqref{eq:third_order_solvable_no_linear} only has three roots if $2^m-1$ is divisible by $3$. 

Solving a first degree polynomial, given by 
\begin{equation}
    \Lambda(X) = 1 + S_1X, 
\end{equation}
is trivial. The polynomial can be rewritten into the form
\begin{equation}
    \Lambda'(X) = \frac{1}{S_1} + X,
    \label{eq:single_error}
\end{equation}
which has the root 
\begin{equation}
    X_1=S_1^{-1}.
    \label{eq:root_single_error}
\end{equation}

\subsection{Solving Second Order Polynomials}
For $t=2$, the general form of the polynomial can be found as~\cite{Polkinghorn1966TIT}
\begin{equation}
    \Lambda(X)=X^2 + S_1X + \frac{S_1^3 + S_3}{S_1}.
\end{equation}
In case $S_1^3 = S_3$, there is a single error and \eqref{eq:single_error} can be used.
In the case where $S_1^3 \neq S_3$ and $S_1\neq 0$, there exist two errors. In this case, by replacing $X=S_1Y$, the polynomial can be transformed into \eqref{eq:second_order_solvable} with $k=(S_1^3+S_3)/S_1^3$. The roots can then be found using 
\begin{equation}
    (X_1,X_2)=S_1\{(S_1^3+S_3)/S_1^3\}_A
    \label{eq:root_double_error}
\end{equation}
An overview of the error conditions is shown in Table~\ref{tab:t2_errors}.

\begin{table}[]
    \centering
    \caption{Error distribution for $t=2$ BCH decoder}
    \label{tab:t2_errors}
    \begin{tabular}{c|c}
     Errors  & Condition \\ 
    \hline
    $0$   & $S_1=S_3=0$ \\
    $1$   & $S_1\neq0$ and $S_1^3=S_3$ \\  
    $2$   & $S_1\neq0$ and $S_1^3\neq S_3$ \\
    $3+$  & $S_1=0$ and $S_3\neq0$ \\
    \end{tabular}
\end{table}

\subsection{Solving Third Order Polynomials}
The general form of the third degree polynomial is~\cite{Polkinghorn1966TIT}
\begin{equation}
\begin{aligned}
    \Lambda(X) = X^3 + S_1X^2 + \frac{S_1^2S_3 + S_5}{S_1^3 + S_3}X \\+  S_1^3 + S_3 + \frac{S_1\left( S_1^2S_3 + S_5\right)}{S_1^3 + S_3}.
    \label{eq:3errs_standard}
\end{aligned}
\end{equation}
For a non-zero syndrome, there exist a total of four configurations of the syndrome components for which an error pattern can be found.
These are shown in Table~\ref{tab:t3_errors}. 
In the case that $S_1^3 = S_3$ and $S_1^5 = S_5$, the received sequence has only a single error, which means that \eqref{eq:single_error} can be used for this first degree polynomial. 
If we have $S_3\left(S_1^3 + S_3 \right) = S_1\left(S_1^5 + S_5\right)$ and $S_1^3\neq S_3$, the received sequence has two errors, in which case the roots can be found using \eqref{eq:root_double_error}.
In case $S_1^3 = S_3$ and $S_1^5 \neq S_5$, the decoder fails. 
If $S_1^5=S_5$ and $S_1^3\neq S_3$, the polynomial can be transformed into \eqref{eq:third_order_solvable_no_linear} with $k=S_1^3+S_3$, if and only if $2^m-1 \text{ mod }3=0$. 
The roots can be found using
\begin{equation}
    (Y_1,Y_2,Y_3) = \{S_1^3 + S_3\}_C,
\end{equation}
with
\begin{equation}
    X_i = Y_i+S_1.
\end{equation}
Otherwise there are no roots.

The last case is when $S_1^3 \neq S_3$ and $S_1^5 \neq S_5$. In this case the polynomial can be transformed into \eqref{eq:third_order_solvable} by using~\cite{Polkinghorn1966TIT}
\begin{equation}
    X = \left( \frac{S_1^5 + S_5}{S_1^3 + S_3}\right)^{-1/2}Z+S_1.
    \label{eq:root_X_triple}
\end{equation}
The roots $Z_1,Z_2, Z_3$ can then be found by using 
\begin{equation}
    (Z_1,Z_2,Z_3) = \left\{(S_1^3 + S_3)^{5/2}(S_1^5 + S_5)^{-3/2}\right\}_B.
    \label{eq:root_Z_triple}
\end{equation}
\begin{table}[]
    \centering
    \caption{Error distribution for $t=3$ BCH decoder}
    \label{tab:t3_errors}
    \begin{tabular}{c|c}
     Errors  & Condition \\
    \hline
    $0$   & $S_1=S_3=S_5=0$\\
    $1$   & $S_1^3=S_3$ and $S_1^5=S_5$\\
    $2$   & $S_3(S_1^3+S_3) = S_1(S_1^5+S_5)$\\
    $3$   & $S_1^3 \neq S_3$ and $S_1^5 \neq S_5$ or\\
          & $S_1^3 \neq S_3$ and $S_1^5 = S_5$ and $2^m-1 \text{ mod } 3 = 0$\\
    $4+$  & $S_1^3 = S_3$ and $S_1^5\neq S_5$ or\\
          & $S_1^3 \neq S_3$ and $S_1^5 = S_5$ and $2^m-1 \text{ mod } 3 \neq 0$
    \end{tabular}
\end{table}

\subsection{Solving Fourth Order Polynomials}
The last direct BCH decoder has $t=4$. The first step here is finding the polynomial. We use the inversionless fourth degree polynomial defined in~\cite[Sec. III-A]{Fougstedt2019JLT}. 
For a given set of syndrome components, the number of errors is determined according to the conditions in Table~\ref{tab:t4_errors}. In case the decoder detects four errors, $\Lambda(X)$ is calculated. Given a particular ELP,  the root-finding process then consists of four different cases (see Table~\ref{tab:solved_cases}). 
Each of these cases finds the roots by first transforming the polynomial into either the form
\begin{equation}
    \Lambda'(Z)=Z^4 + Z^2 + k_1Z + k_2
    \label{eq:poly_4_v1}
\end{equation}
or into the form
\begin{equation}
    \Lambda'(Z)=Z^4 + k_1Z + k_2.
    \label{eq:poly_4_v2}
\end{equation}
For both these forms the roots can then be found using a small number of computations. For \eqref{eq:poly_4_v1}, the first root can be computed using~\cite[eq. (16)]{Yan1998EL}
\begin{equation}
    Z_1 = b_1\left\{(1+b_1^{-2})\left\{\frac{k_2}{1+b_1^4}\right\}_A\right\}_A,
    \label{eq:Z1_v1}
\end{equation}
and for~\eqref{eq:poly_4_v2}
\begin{equation}
    Z_1 = c_1\left\{\left\{\frac{k_2}{c_1^{4}}\right\}_A\right\}_A.
    \label{eq:Z1_v2}
\end{equation}
The parameters $b_i$ and $c_i$ refer to the $i$th root of \eqref{eq:third_order_solvable} and \eqref{eq:third_order_solvable_no_linear}, respectively. 
To find the variables $k_1$ and $k_2$ there are four options~\cite{Yan1998EL}. 
In the first two cases in Table~\ref{tab:solved_cases}, $k_1$ and $k_2$ can be directly defined based on the coefficients of $\Lambda(X)$. The change of variables is then by $X=Z$ for the top row and
\begin{equation}
    X=Z\left(\frac{\Lambda_2}{\Lambda_4}\right)^{1/2}.
    \label{eq:x_z_case2}
\end{equation}
For the other two cases, the polynomial is first transformed by using
\begin{equation}
    X = 1/Y + \left(\frac{\Lambda_1}{\Lambda_3}\right)^{1/2},
    \label{eq:fourth_order_v1}
\end{equation}
giving the polynomial,
\begin{equation}
    q(Y) = q_4Y^4 + q_2Y^2 + q_1Y + 1,
\end{equation}
where 
\begin{equation}
    q_4 = \frac{\Lambda_0}{\Lambda_4} + \frac{\Lambda_1\Lambda_2}{\Lambda_3\Lambda_4} + \left(\frac{\Lambda_1}{\Lambda_3}\right)^2,
\end{equation}
and
\begin{equation}
    q_2 = \frac{\Lambda_2}{\Lambda_4} + \frac{\left(\Lambda_1\Lambda_3\right)^{1/2}}{\Lambda_4},
\end{equation}
and $q_1 = \Lambda_3/\Lambda_1$.
If $q_2=0$ this polynomial can be transformed into \eqref{eq:poly_4_v2} with $k_1=q_1/q_4$ and $k_2=1/q_4$. 
If $q_2\neq0$, the polynomial can be transformed into \eqref{eq:poly_4_v1} by taking $Y=Z(q_2/q_4)^{1/2}$, which gives
\begin{equation}
    X = \frac{1}{Z}\left(\frac{q_4}{q_2}\right)^{1/2}+ \left(\frac{\Lambda_1}{\Lambda_3}\right)^{1/2}.
    \label{eq:fourth_order_v2}
\end{equation}
This transformation then gives $k_1=q_1(q_4/q_2^3)^{1/2}$ and $k_2=q_4/q_2^2$.
When root $Z_1$ has been found, the remaining roots can be found according to $Z_2=Z_1+b_1$ or $Z_2=Z_1+c_1$, $Z_3=Z_1+b_2$ or $Z_3=Z_1+c_2$, and $Z_4=Z_1+Z_2+Z_3$. The roots $X_i$ can then be calculated using the equations in column $X$ in Table~\ref{tab:solved_cases}.

\begin{table}[]
    \centering
    \caption{Error distribution for $t=4$ BCH decoder}
    \label{tab:t4_errors}
    {\renewcommand{\arraystretch}{1.5}%
    \begin{tabular}{c|c}
     Errors  & Condition \\
    \hline
    $0$   & $S_1=S_3=S_5=S_7=0$\\
    \hline
    $1$   & $S_1^3=S_3$ and $S_1^5=S_5$ and $S_1^7=S_7$\\
    \hline
    $2$   & $S_3(S_1^3+S_3) = S_1(S_1^5+S_5)$ and $S_1\neq0$ and \\
          & $S_1^7  + S_1^2S_3^2=S_5(S_1^3+S_3)$\\
    \hline
    \multirow{3}{*}{$3$}   & $(S_1S_7+S_1^2S_3^2 + S_3(S_1^5+S_5))(S^3_1+S_3)=$\\
                           & $S_5(S_1^3+S_3)^2 + S_1(S_1^5+S_5)^2$ and $S_1^3\neq S_3$ and $S_1\neq0$\\
    \cline{2-2}
          & or $S_3S_7 = S_5^2$ and $S_1=0$\\
    \hline
    $4+$  & Otherwise\\
    \end{tabular}}
\end{table}

\begin{table}
    \centering
    \caption{Parameter cases for fourth order polynomials}
    \label{tab:solved_cases}
    \begin{tabular}{c|c|c|c|c|c|c}
        $\Lambda_2$ & $\Lambda_3$ & $q_2$ & $X$ & eq. & $k_1$ & $k_2$\\
        \hline
        $0$     &  $0$      & - & $X=Z$ & \eqref{eq:Z1_v2}  & $\Lambda_1/\Lambda_4$ & $\Lambda_0/\Lambda_4$\\
        $\neq0$     & $0$   & - & \eqref{eq:x_z_case2}& \eqref{eq:Z1_v1} & $\Lambda_1\Lambda_4^{1/2}/\Lambda_2^{3/2}$ & $\Lambda_0\Lambda_4/\Lambda_2^2$\\
        
        $\neq0$ & $\neq0$ & $0$ & \eqref{eq:fourth_order_v1} & \eqref{eq:Z1_v2} & $q_1/q_4$ & $1/q_4$ \\ 
        
        $\neq0$ & $\neq0$ & $\neq0$ & \eqref{eq:fourth_order_v2} & \eqref{eq:Z1_v1} & $q_1(q_4/q_2^3)^{1/2}$ & $q_4/q_2^2$\\ 

    \end{tabular}
\end{table}

%% file: Sections/hardware_implementation.tex
\section{Hardware Implementation}
\label{hardware_architecture}
In this section we discuss a set of general GF($2$) computation functions that are shared between the two decoders. Next, we discuss our hardware architecture for the conventional decoder, followed by the architecture for the direct decoder. 

\subsection{GF(2) Computations}
For the decoding of BCH codes we consider four types of computations: multiplication, division, raising to an integer power and raising to a non-integer power, i.e., square root or cube root. Multiplication of two Galois elements $\alpha^i$ and $\alpha^j$, can be achieved with only \texttt{xor} and \texttt{and} gates by unrolling the multiplication. 
Similarly, raising an element $\alpha^i$ to a power can be unrolled, with the added benefit that, in GF(2), a large number of gates can be optimized away. 
Division of two elements $\alpha^i/\alpha^j$ is equivalent to multiplying $\alpha^i$ with the inverse $\alpha^{-j}$. As there are $2^m$ elements for a field with power $m$, and for each element there exists exactly one unique inverse, inversion is achieved through a LUT. Division is implemented by combining an inversion LUT with multiplication. 
Similarly, computing square roots and cube roots is achieved with a LUT.  
Computing the syndrome components is achieved through unrolled matrix multiplication, as the parity matrix is known at compile time. 

\subsection{Berlekamp-Massey and Chien Search}
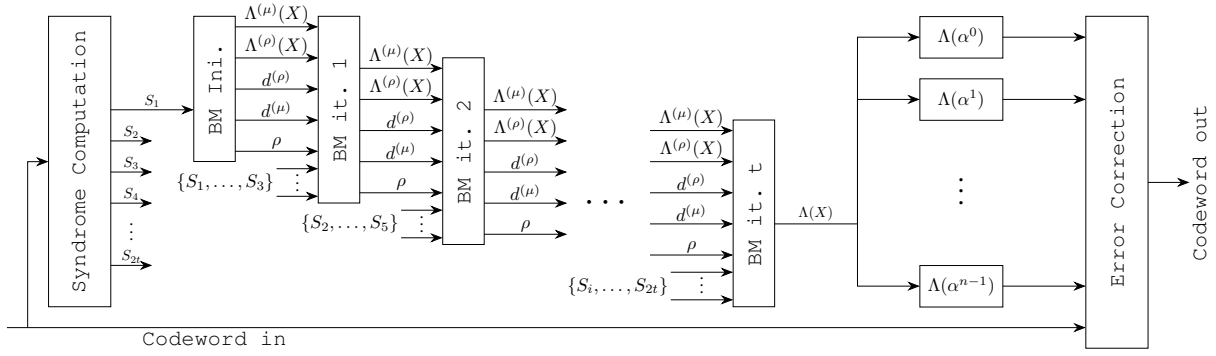
\begin{figure*}[]
    \centering
    \resizebox{1.8\columnwidth}{!}{\input{Figures/BCH_bm_chien.tikz}}
    \caption{Block diagram of the BM and Chien-search BCH decoder implemented in this work. The architecture consists of a syndrome computation block, $t$ iterations of the BM update, a Chien-search block, and a final error-correction stage.}
    \label{fig:BM-Chien}
\end{figure*}

The conventional implementation consists of an initialization step of one clock cycle, $t$ iterations of two clock cycles for the BM steps, and a single clock cycle to compute the Chien search. A block diagram is shown in Fig.~\ref{fig:BM-Chien}. In the first step, the syndrome components $S_1,\dots,S_{2t}$ are calculated and BM initialization is applied (lines~\ref{alg_line:init_start}-\ref{alg_line:init_end} in Alg.~\ref{alg:berlekamp_massey}). This step is achieved in a single clock cycle. During the $t$ BM iterations, the architecture computes $\Lambda^{(\mu)}(X)$ in the first clock cycle and computes $d^{(\mu)}$ and $d^{(\rho)}$ in the second, with BM requiring a total of $2t$ clock cycles. After the BM has successfully computed an ELP, Chien search is implemented in a fully parallel manner. 
The latency of our conventional decoder is given by $2t+2$ clock cycles.

\subsection{Direct Implementation}
\begin{figure*}
    \centering
    \resizebox{1.8\columnwidth}{!}{\input{Figures/BCH_direct.tikz}}
    \caption{Block diagram of the $t=4$ direct decoder implemented in this work. The architecture consists of a syndrome computation block, a precomputation block, a determine errors block, multiple root computation stages and an error-correction stage.}
    \label{fig:direct_decoder}
\end{figure*}
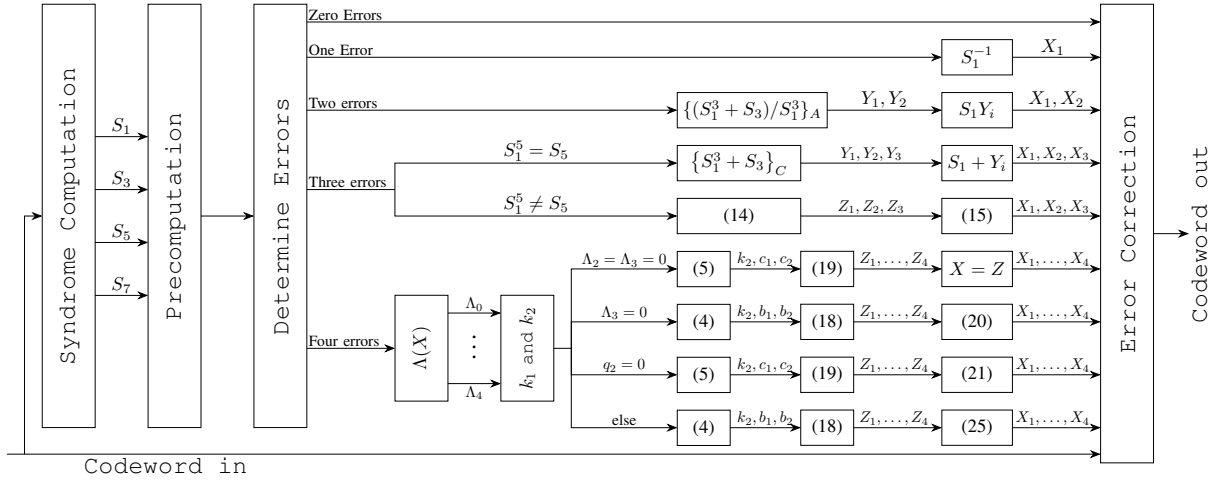

The direct implementations are divided into five parts: (1) computing the syndrome components, (2) pre-computations with the syndrome components, (3) computing the number of errors, (4) computing the roots, and (5) correcting the errors. 
Computing the syndrome components is achieved in the same manner as the conventional implementation. 
Fig.~\ref{fig:direct_decoder} shows the block diagram for a direct decoder with $t=4$. 
The \texttt{determine errors} block in Fig.~\ref{fig:direct_decoder} selects the correct state based on the parameters in Table~\ref{tab:t4_errors}. 
The decoders for $t=1$, $t=2$, and $t=3$ are similar, with changes to the \texttt{pre-computation} and \texttt{determine errors} blocks and the fact that the correction of more than $t$ errors is absent. 
Our direct decoders have a latency of $3$, $4$, and $8$ clock cycles for $t=2$, $t=3$ and $t=4$, respectively.

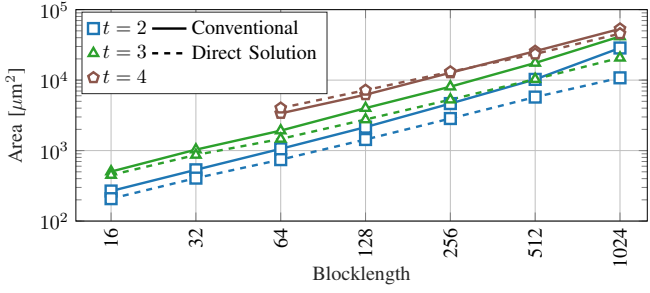
\begin{figure}
    \centering
    \resizebox{8cm}{!}{\adjustbox{trim=20mm 0mm 0mm 0mm}{\input{Figures/ASIC_area.tikz}}}
    \caption{Post-synthesis area for 16~nm FinFET at $1$ GHz.}
    \label{fig:ASIC_area}
\end{figure}

%% file: Figures/ASIC_area.tikz
\definecolor{Cblue}{RGB}{31 119 180}
\definecolor{Cred}{RGB}{214 39 40}
\definecolor{Corange}{RGB}{255 127 14}
\definecolor{Cgreen}{RGB}{44 160 44}
\definecolor{Cpurple}{RGB}{148 103 198}
\definecolor{Cbrown}{RGB}{140 86 75}
\definecolor{Cpink}{RGB}{227 119 194}
\definecolor{Ccyan}{RGB}{23 190 207}
\definecolor{Cgray}{RGB}{127 127 127}
\definecolor{Cyellow}{RGB}{188 189 34}

\begin{tikzpicture}

\begin{axis}[%
width=10cm,
height=3.7cm,
at={(0, 0)},
scale only axis,
xmode=log,
xmin=12,
xmax=1280,
xtick={8, 16, 32, 64, 128, 256, 512, 1024},
xticklabels={8, 16, 32, 64, 128, 256, 512, 1024}, 
x tick label style={rotate=90,anchor=east},
ymode=log,
ymin=100,
ymax=100000,
yminorticks=true,
axis background/.style={fill=white},
xmajorgrids,
xminorticks=true,
ymajorgrids,
yminorgrids=false,
xlabel style={font=\color{white!15!black}, at={(axis description cs:0.5,-.18)},anchor=north},
xlabel={Blocklength},  
ylabel style={font=\color{white!15!black}},
ylabel={Area [$\mu \text{m}^2$]},
legend columns=2,
legend style={at={(0.01, 0.99)}, anchor=north west, legend cell align=left, align=left, draw=white!15!black}
]

\addlegendimage{color=Cblue, only marks, solid,  mark=square*, mark options={solid, Cblue, fill=white, scale=1.35}, very thick}
\addlegendentry{$t=2$}

\addlegendimage{color=black, solid, very thick}
\addlegendentry{Conventional}

\addlegendimage{color=Cgreen, only marks, solid,  mark=triangle*, mark options={solid, Cgreen, fill=white, scale=1.35}, very thick}
\addlegendentry{$t=3$}

\addlegendimage{color=black, dashed, very thick}
\addlegendentry{Direct Solution}

\addlegendimage{color=Cbrown, only marks, solid,  mark=pentagon*, mark options={solid, Cbrown, fill=white, scale=1.35}, very thick}
\addlegendentry{$t=4$}

\addplot [color=Cblue, solid,  mark=square*, mark options={solid, Cblue, fill=white, scale=1.35}, very thick]
  table[row sep=crcr]{%
  16 266.92 \\
  32 533.9001 \\
  64 1059.4022 \\
  128 2152.35 \\
  256 4648.28 \\
  512 10242.344 \\
  1024 28556.271899 \\
};
	
\addplot [color=Cblue, dashed,  mark=square*, mark options={solid, Cblue, fill=white, scale=1.35}, very thick]
  table[row sep=crcr]{%
    16 207.878404   \\
    32 406.840327   \\
    64 744.370572   \\
    128 1437.108501 \\
    256 2851.096362 \\
    512 5745.945692 \\
    1024 10777.536164 \\
};

\addplot [color=Cgreen, solid,  mark=triangle*, mark options={solid, Cgreen, fill=white, scale=1.35}, very thick]
  table[row sep=crcr]{%
  16 503.47 \\
  32 1031.512 \\
  64 1921.2122 \\
  128 4004.69 \\
  256 8067.65 \\
  512 17509.06 \\
  1024 41600.512243 \\
};
	
\addplot [color=Cgreen, dashed,  mark=triangle*, mark options={solid, Cgreen, fill=white, scale=1.35}, very thick]
  table[row sep=crcr]{%
  16 449.919\\
  32 867.075 \\
  64 1463.7024 \\
  128 2760.376 \\
  256 5277.571 \\
  512 10325.802 \\
  1024 20716.093 \\
};

\addplot [color=Cbrown, solid,  mark=pentagon*, mark options={solid, Cbrown, fill=white, scale=1.35}, very thick]
  table[row sep=crcr]{%
  64 3377.116855 \\
  128 6211.365183 \\
  256 12714 \\
  512 25706.004842 \\
  1024 53293.024502 \\
};
	
\addplot [color=Cbrown, dashed,  mark=pentagon*, mark options={solid, Cbrown, fill=white, scale=1.35}, very thick]
  table[row sep=crcr]{%
    64 4018.273999 \\
    128 7195.806847 \\ 
    256 13118 \\
    512 23485.334780 \\ 
    1024 45605.618747 \\
};

\end{axis}
\end{tikzpicture}%

%% file: Figures/BCH_bm_chien.tikz
\begin{tikzpicture}

    \node[anchor=north west, draw, minimum width=14cm, minimum height = 3cm, rotate=90] at (0, -2) {\scalebox{3}{\texttt{Syndrome Computation}}};

    \node[anchor=north west, draw, minimum width=7cm, minimum height = 2cm, rotate=90] at (7, 5) {\scalebox{3}{\texttt{BM Ini.}}};
    
    \draw[-{Stealth[scale=2]}, very thick] (9, 11.5) -- node[above, anchor=south] {\scalebox{2.5}{$\Lambda^{(\mu)}(X)$}}  (13, 11.5); 
    \draw[-{Stealth[scale=2]}, very thick] (9, 10) -- node[above, anchor=south] {\scalebox{2.5}{$\Lambda^{(\rho)}(X)$}}  (13, 10); 
    \draw[-{Stealth[scale=2]}, very thick] (9, 8.5) -- node[above, anchor=south] {\scalebox{2.5}{$d^{(\rho)}$}}  (13, 8.5); 
    \draw[-{Stealth[scale=2]}, very thick] (9, 7) -- node[above, anchor=south] {\scalebox{2.5}{$d^{(\mu)}$}}  (13, 7); 
    \draw[-{Stealth[scale=2]}, very thick] (9, 5.5) -- node[above, anchor=south] {\scalebox{2.5}{$\rho$}}  (13, 5.5); 

        \draw[-{Stealth[scale=2]}, very thick] (3, 7.5) -- node[above, anchor=south] {\scalebox{2}{$S_1$}}  (7, 7.5);     
        \draw[-{Stealth[scale=2]}, very thick] (3, 6) -- node[above, anchor=south] {\scalebox{2}{$S_2$}}  (5, 6);
        \draw[-{Stealth[scale=2]}, very thick] (3, 4.5) -- node[above, anchor=south] {\scalebox{2}{$S_3$}}  (5, 4.5);
        \draw[-{Stealth[scale=2]}, very thick] (3, 3) -- node[above, anchor=south] {\scalebox{2}{$S_4$}}  (5, 3); 
        \node at (4, 1.875) {\scalebox{2.5}{$\vdots$}};
        \draw[-{Stealth[scale=2]}, very thick] (3, 0) -- node[above, anchor=south] {\scalebox{2}{$S_{2t}$}}  (5, 0);
        
        \node[anchor=north west, draw, minimum width=9cm, minimum height = 2cm, rotate=90] at (13, 3) {\scalebox{3}{\texttt{BM it. 1}}};
        \draw[-{Stealth[scale=2]}, very thick] (11, 4.66) -- (13, 4.66); 
        \node[anchor=center] at (12, 4.25) {\scalebox{2.5}{$\vdots$}};
        \draw[-{Stealth[scale=2]}, very thick] (11, 3.33) -- (13, 3.33); 
        \node[anchor=east] at (11, 4) {\scalebox{2.5}{$\{S_1,\dots,S_3\}$}};

        \draw[-{Stealth[scale=2]}, very thick] (15, 9.5) -- node[above, anchor=south] {\scalebox{2.5}{$\Lambda^{(\mu)}(X)$}}  (19, 9.5); 
        \draw[-{Stealth[scale=2]}, very thick] (15, 8) -- node[above, anchor=south] {\scalebox{2.5}{$\Lambda^{(\rho)}(X)$}}  (19, 8); 
        \draw[-{Stealth[scale=2]}, very thick] (15, 6.5) -- node[above, anchor=south] {\scalebox{2.5}{$d^{(\rho)}$}}  (19, 6.5); 
        \draw[-{Stealth[scale=2]}, very thick] (15, 5) -- node[above, anchor=south] {\scalebox{2.5}{$d^{(\mu)}$}}  (19, 5); 
        \draw[-{Stealth[scale=2]}, very thick] (15, 3.5) -- node[above, anchor=south] {\scalebox{2.5}{$\rho$}}  (19, 3.5); 

        \node[anchor=north west, draw, minimum width=9cm, minimum height = 2cm, rotate=90] at (19, 1) {\scalebox{3}{\texttt{BM it. 2}}};

        \draw[-{Stealth[scale=2]}, very thick] (17, 2.66) -- (19, 2.66); 
        \node[anchor=center] at (18, 2.25) {\scalebox{2.5}{$\vdots$}};
        \draw[-{Stealth[scale=2]}, very thick] (17, 1.33) -- (19, 1.33); 
        \node[anchor=east] at (17, 2) {\scalebox{2.5}{$\{S_2,\dots,S_5\}$}};

        \draw[-{Stealth[scale=2]}, very thick] (21, 7.5) -- node[above, anchor=south] {\scalebox{2.5}{$\Lambda^{(\mu)}(X)$}}  (25, 7.5); 
        \draw[-{Stealth[scale=2]}, very thick] (21, 6) -- node[above, anchor=south] {\scalebox{2.5}{$\Lambda^{(\rho)}(X)$}}  (25, 6); 
        \draw[-{Stealth[scale=2]}, very thick] (21, 4.5) -- node[above, anchor=south] {\scalebox{2.5}{$d^{(\rho)}$}}  (25, 4.5); 
        \draw[-{Stealth[scale=2]}, very thick] (21, 3) -- node[above, anchor=south] {\scalebox{2.5}{$d^{(\mu)}$}}  (25, 3); 
        \draw[-{Stealth[scale=2]}, very thick] (21, 1.5) -- node[above, anchor=south] {\scalebox{2.5}{$\rho$}}  (25, 1.5); 
         

        \pgfmathtruncatemacro\FinalItStart{30}

        \node[anchor=north west, draw, minimum width=9cm, minimum height = 2cm, rotate=90] at (\FinalItStart+3, -2) {\scalebox{3}{\texttt{BM it. t}}};

        \draw[-{Stealth[scale=2]}, very thick] (\FinalItStart-1, 6.5) -- node[above, anchor=south] {\scalebox{2.5}{$\Lambda^{(\mu)}(X)$}}  (\FinalItStart+3, 6.5); 
        \draw[-{Stealth[scale=2]}, very thick] (\FinalItStart-1, 5) -- node[above, anchor=south] {\scalebox{2.5}{$\Lambda^{(\rho)}(X)$}}  (\FinalItStart+3, 5); 
        \draw[-{Stealth[scale=2]}, very thick] (\FinalItStart-1, 3.5) -- node[above, anchor=south] {\scalebox{2.5}{$d^{(\rho)}$}}  (\FinalItStart+3, 3.5); 
        \draw[-{Stealth[scale=2]}, very thick] (\FinalItStart-1, 2) -- node[above, anchor=south] {\scalebox{2.5}{$d^{(\mu)}$}}  (\FinalItStart+3, 2); 
        \draw[-{Stealth[scale=2]}, very thick] (\FinalItStart-1, 0.5) -- node[above, anchor=south] {\scalebox{2.5}{$\rho$}}  (\FinalItStart+3, 0.5);

        \draw[-{Stealth[scale=2]}, very thick] (\FinalItStart, -0.33) -- (\FinalItStart+3, -0.33); 
        \node at (\FinalItStart+1.5, -0.7) {\scalebox{2.5}{$\vdots$}};
        \draw[-{Stealth[scale=2]}, very thick] (\FinalItStart, -1.66) -- (\FinalItStart+3, -1.66); 
        \node[anchor=east] at (\FinalItStart, -1) {\scalebox{2.5}{$\{S_i,\dots,S_{2t}\}$}};

        \draw[-, very thick] (\FinalItStart+5, 2) -- node[above, anchor=south] {\scalebox{2}{$\Lambda(X)$}} (\FinalItStart+9, 2);
        
        \node[] at (\FinalItStart-3, 3) {\scalebox{5}{$\cdots$}};
    
        \pgfmathtruncatemacro\ChienStart{42}
        \node[anchor=north west, draw, minimum width=4cm, minimum height = 2cm, rotate=0] at (\ChienStart, 12) {\scalebox{2.5}{\texttt{$\Lambda(\alpha^0)$}}};
        \node[anchor=north west, draw, minimum width=4cm, minimum height = 2cm, rotate=0] at (\ChienStart, 9) {\scalebox{2.5}{\texttt{$\Lambda(\alpha^1)$}}};        
        
        \node[] at (\ChienStart+2, 4) {\scalebox{4}{$\vdots$}};

        \node[anchor=north west, draw, minimum width=4cm, minimum height = 2cm, rotate=0] at (\ChienStart, 0) {\scalebox{2.5}{\texttt{$\Lambda(\alpha^{n-1})$}}};

        \draw[-{Stealth[scale=2]}, very thick] (\FinalItStart+9, 2) -- (\FinalItStart+9, 11) -- (\ChienStart, 11);
        \draw[-{Stealth[scale=2]}, very thick] (\FinalItStart+9, 2) -- (\FinalItStart+9, 8) -- (\ChienStart, 8);
        \draw[-{Stealth[scale=2]}, very thick] (\FinalItStart+9, 2) -- (\FinalItStart+9, -1) -- (\ChienStart, -1);

        \pgfmathtruncatemacro\ECCStart{50}

        \node[anchor=north west, draw, minimum width=16cm, minimum height = 3cm, rotate=90] at (\ECCStart, -4) {\scalebox{3}{\texttt{Error Correction}}};

        \draw[-{Stealth[scale=2]}, very thick] (\ChienStart+4, 11) -- (\ECCStart, 11);
        \draw[-{Stealth[scale=2]}, very thick] (\ChienStart+4, 8) -- (\ECCStart, 8);
        \draw[-{Stealth[scale=2]}, very thick] (\ChienStart+4, -1) -- (\ECCStart, -1);
        \draw[-{Stealth[scale=2]}, very thick] (-2, -3) -- node[below, anchor=north, xshift=-16cm] {\scalebox{3}{\texttt{Codeword in}}} (\ECCStart, -3);
        \draw[-{Stealth[scale=2]}, very thick] (-1, -3) -- (-1, 5) -- (0, 5); 

        \draw[-{Stealth[scale=2]}, very thick] (\ECCStart+3, 4) -- node[below, anchor=north, rotate=90, yshift=-1cm] {\scalebox{3}{\texttt{Codeword out}}} (\ECCStart+5, 4); 
\end{tikzpicture}

%% file: Figures/BCH_direct.tikz
\begin{tikzpicture}

    \node[anchor=north west, draw, minimum width=24cm, minimum height = 3cm, rotate=90] at (0, -12) {\scalebox{4}{\texttt{Syndrome Computation}}};

    \pgfmathtruncatemacro\PrecompStart{6}
    \pgfmathtruncatemacro\DetErrsStart{12}
    \pgfmathtruncatemacro\ECCStart{60}
    \node[anchor=north west, draw, minimum width=24cm, minimum height = 3cm, rotate=90] at (\PrecompStart, -12) {\scalebox{4}{\texttt{Precomputation}}};

    \draw[-{Stealth[scale=2]}, very thick] (\PrecompStart+3, 0) -- (\DetErrsStart, 0); 

    \draw[-{Stealth[scale=2]}, very thick] (3, 4.5) -- node[above, anchor=south] {\scalebox{3}{$S_1$}} (\PrecompStart, 4.5);
    \draw[-{Stealth[scale=2]}, very thick] (3, 1.5) -- node[above, anchor=south] {\scalebox{3}{$S_3$}} (\PrecompStart, 1.5);
    \draw[-{Stealth[scale=2]}, very thick] (3, -1.5) -- node[above, anchor=south] {\scalebox{3}{$S_5$}} (\PrecompStart, -1.5);
    \draw[-{Stealth[scale=2]}, very thick] (3, -4.5) -- node[above, anchor=south] {\scalebox{3}{$S_7$}} (\PrecompStart, -4.5);
    
    \node[anchor=north west, draw, minimum width=24cm, minimum height = 3cm, rotate=90] at (\DetErrsStart, -12) {\scalebox{4}{\texttt{Determine Errors}}};

        \draw[-{Stealth[scale=2]}, very thick] (\DetErrsStart+3, 11) -- (\ECCStart, 11); 
        \node[anchor=south west] at (\DetErrsStart+3, 11) {\scalebox{2.5}{Zero Errors}}; 

        \draw[-{Stealth[scale=2]}, very thick] (\DetErrsStart+3, 9) -- (\ECCStart-9, 9);
        \node[anchor=west, draw, minimum width=4cm, minimum height=2cm] at (\ECCStart-9, 9) {\scalebox{3}{$S_1^{-1}$}};
        \draw[-{Stealth[scale=2]}, very thick] (\ECCStart-5, 9) --  node[above, anchor=south, xshift=-2mm] {\scalebox{3}{$X_1$}} (\ECCStart, 9);
        \node[anchor=south west] at (\DetErrsStart+3, 9) {\scalebox{2.5}{One Error}}; 

        \draw[-{Stealth[scale=2]}, very thick] (\DetErrsStart+3, 6) -- (\ECCStart-24, 6);
        \node[anchor=west, draw, minimum width=8.5cm, minimum height=2cm] at (\ECCStart-24, 6) {\scalebox{3}{$\{(S_1^3+S_3)/S_1^3\}_A$}};
        \node[anchor=west, draw, minimum width=4cm, minimum height=2cm] at (\ECCStart-9, 6) {\scalebox{3}{$S_1Y_i$}};
        \draw[-{Stealth[scale=2]}, very thick] (\ECCStart-15.5, 6) -- node[above, anchor=south] {\scalebox{3}{$Y_1,Y_2$}}(\ECCStart-9, 6);         
        \draw[-{Stealth[scale=2]}, very thick] (\ECCStart-5, 6) --  node[above, anchor=south] {\scalebox{3}{$X_1,X_2$}} (\ECCStart, 6);
        \node[anchor=south west] at (\DetErrsStart+3, 6) {\scalebox{2.5}{Two errors}}; 

        \node[anchor=west, draw, minimum width=7cm, minimum height=2cm] at (\ECCStart-24, 3) {\scalebox{3}{$\left\{S_1^3+S_3\right\}_C$}};
        \node[anchor=west, draw, minimum width=7cm, minimum height=2cm] at (\ECCStart-24, 0) {\scalebox{3}{\eqref{eq:root_X_triple}}};

        \draw[-{Stealth[scale=2]}, very thick] (\ECCStart-17, 3) -- node[above, anchor=south] {\scalebox{2.5}{$Y_1,Y_2,Y_3$}} (\ECCStart-9, 3);
        \draw[-{Stealth[scale=2]}, very thick] (\ECCStart-17, 0) -- node[above, anchor=south] {\scalebox{2.5}{$Z_1,Z_2,Z_3$}} (\ECCStart-9, 0);
        
        \node[anchor=west, draw, minimum width=4cm, minimum height=2cm] at (\ECCStart-9, 3) {\scalebox{3}{$S_1+Y_i$}};
        \node[anchor=west, draw, minimum width=4cm, minimum height=2cm] at (\ECCStart-9, 0) {\scalebox{3}{\eqref{eq:root_Z_triple}}};

        \draw[-{Stealth[scale=2]}, very thick] (\ECCStart-5, 3) -- node[above, anchor=south, xshift=-2mm] {\scalebox{2.5}{$X_1,X_2,X_3$}} (\ECCStart, 3);
        \draw[-{Stealth[scale=2]}, very thick] (\ECCStart-5, 0) -- node[above, anchor=south, xshift=-2mm] {\scalebox{2.5}{$X_1,X_2,X_3$}} (\ECCStart, 0);

        \draw[-, very thick] (\DetErrsStart+3, 1.5) -- (\DetErrsStart+8, 1.5); 
        \draw[-{Stealth[scale=2]}, very thick] (\DetErrsStart+8, 1.5) -- (\DetErrsStart+8, 3) -- node[above, anchor=south] {\scalebox{3}{$S_1^5=S_5$}} (\ECCStart-24, 3); 
        \draw[-{Stealth[scale=2]}, very thick] (\DetErrsStart+8, 1.5) -- (\DetErrsStart+8, 0) -- node[above, anchor=south] {\scalebox{3}{$S_1^5\neq S_5$}} (\ECCStart-24, 0); 
        \node[anchor=south west] at (\DetErrsStart+3, 1.5) {\scalebox{2.5}{Three errors}}; 
    
        \node[anchor=south west] at (\DetErrsStart+3, -7.5) {\scalebox{2.5}{Four errors}}; 
        \draw[-{Stealth[scale=2]}, very thick] (\DetErrsStart+3, -7.5) -- (\DetErrsStart+8, -7.5); 
        \node[anchor=north west, draw, minimum width=6cm, minimum height = 3cm, rotate=90] at (\DetErrsStart+8, -10.5) {\scalebox{3}{$\Lambda(X)$}};
        \draw[-{Stealth[scale=2]}, very thick] (\DetErrsStart+11, -5.5) -- node[above, anchor=south] {\scalebox{2.5}{$\Lambda_0$}}(\DetErrsStart+14, -5.5);
        \node at (\DetErrsStart+12.5, -7) {\scalebox{4}{$\vdots$}};

        \draw[-{Stealth[scale=2]}, very thick] (\DetErrsStart+11, -9.5) -- node[below, anchor=north] {\scalebox{2.5}{$\Lambda_4$}}(\DetErrsStart+14, -9.5);

        \node[anchor=north west, draw, minimum width=6cm, minimum height = 3cm, rotate=90] at (\DetErrsStart+14, -10.5) {\scalebox{3}{$k_1$ \texttt{and} $k_2$}};
        \draw[-, very thick] (\DetErrsStart+17, -7.5) -- (\DetErrsStart+18, -7.5); 

        \draw[-{Stealth[scale=2]}, very thick] (\DetErrsStart+18, -7.5) -- (\DetErrsStart+18, -3) -- node[above, anchor=south] {\scalebox{2.5}{$\Lambda_2=\Lambda_3=0$}} (\DetErrsStart+24, -3);

        \draw[-{Stealth[scale=2]}, very thick] (\DetErrsStart+18, -7.5) -- (\DetErrsStart+18, -6) -- node[above, anchor=south] {\scalebox{2.5}{$\Lambda_3=0$}} (\DetErrsStart+24, -6);

        \draw[-{Stealth[scale=2]}, very thick] (\DetErrsStart+18, -7.5) -- (\DetErrsStart+18, -9) -- node[above, anchor=south] {\scalebox{2.5}{$q_2=0$}} (\DetErrsStart+24, -9);

        \draw[-{Stealth[scale=2]}, very thick] (\DetErrsStart+18, -7.5) -- (\DetErrsStart+18, -12) -- node[above, anchor=south] {\scalebox{2.5}{else}} (\DetErrsStart+24, -12);
        
        \node[anchor=north west, draw, minimum width=3cm, minimum height = 2cm ] at (\DetErrsStart+24, -2) {\scalebox{3}{\eqref{eq:third_order_solvable_no_linear}}};
        \node[anchor=north west, draw, minimum width=3cm, minimum height = 2cm ] at (\DetErrsStart+24, -5) {\scalebox{3}{\eqref{eq:third_order_solvable}}};
        
        \node[anchor=north west, draw, minimum width=3cm, minimum height = 2cm ] at (\DetErrsStart+24, -8) {\scalebox{3}{\eqref{eq:third_order_solvable_no_linear}}};
        \node[anchor=north west, draw, minimum width=3cm, minimum height = 2cm ] at (\DetErrsStart+24, -11) {\scalebox{3}{\eqref{eq:third_order_solvable}}};

        \draw[-{Stealth[scale=2]}, very thick] (\DetErrsStart+27, -3) -- node[above, anchor=south] {\scalebox{2.5}{$k_2,c_1,c_2$}} (\DetErrsStart+31, -3); 

        \draw[-{Stealth[scale=2]}, very thick] (\DetErrsStart+27, -6) -- node[above, anchor=south] {\scalebox{2.5}{$k_2,b_1,b_2$}} (\DetErrsStart+31, -6);

        \draw[-{Stealth[scale=2]}, very thick] (\DetErrsStart+27, -9) -- node[above, anchor=south] {\scalebox{2.5}{$k_2,c_1,c_2$}} (\DetErrsStart+31, -9);

        \draw[-{Stealth[scale=2]}, very thick] (\DetErrsStart+27, -12) -- node[above, anchor=south] {\scalebox{2.5}{$k_2,b_1,b_2$}} (\DetErrsStart+31, -12);

        \node[anchor=west, draw, minimum width=3cm, minimum height=2cm] at ((\DetErrsStart+31, -3) {\scalebox{3}{\eqref{eq:Z1_v2}}}; 

        \node[anchor=west, draw, minimum width=3cm, minimum height=2cm] at ((\DetErrsStart+31, -6) {\scalebox{3}{\eqref{eq:Z1_v1}}}; 

        \node[anchor=west, draw, minimum width=3cm, minimum height=2cm] at ((\DetErrsStart+31, -9) {\scalebox{3}{\eqref{eq:Z1_v2}}};

        \node[anchor=west, draw, minimum width=3cm, minimum height=2cm] at ((\DetErrsStart+31, -12) {\scalebox{3}{\eqref{eq:Z1_v1}}}; 

        \draw[-{Stealth[scale=2]}, very thick] (\DetErrsStart+34, -3) -- node[above, anchor=south, xshift=-2mm] {\scalebox{2.5}{$Z_1,\dots,Z_4$}} (\ECCStart-9, -3);

        \draw[-{Stealth[scale=2]}, very thick] (\DetErrsStart+34, -6) -- node[above, anchor=south, xshift=-2mm] {\scalebox{2.5}{$Z_1,\dots,Z_4$}} (\ECCStart-9, -6);

        \draw[-{Stealth[scale=2]}, very thick] (\DetErrsStart+34, -9) -- node[above, anchor=south, xshift=-2mm] {\scalebox{2.5}{$Z_1,\dots,Z_4$}} (\ECCStart-9, -9);

        \draw[-{Stealth[scale=2]}, very thick] (\DetErrsStart+34, -12) -- node[above, anchor=south, xshift=-2mm] {\scalebox{2.5}{$Z_1,\dots,Z_4$}} (\ECCStart-9, -12);

        \node[anchor=west, draw, minimum width=4cm, minimum height=2cm] at (\ECCStart-9, -3) {\scalebox{3}{$X=Z$}};

        \node[anchor=west, draw, minimum width=4cm, minimum height=2cm] at (\ECCStart-9, -6) {\scalebox{3}{\eqref{eq:x_z_case2}}};

        \node[anchor=west, draw, minimum width=4cm, minimum height=2cm] at (\ECCStart-9, -9) {\scalebox{3}{\eqref{eq:fourth_order_v1}}};

        \node[anchor=west, draw, minimum width=4cm, minimum height=2cm] at (\ECCStart-9, -12) {\scalebox{3}{\eqref{eq:fourth_order_v2}}};

        \draw[-{Stealth[scale=2]}, very thick] (\ECCStart-5, -3) -- node[above, anchor=south, xshift=-2mm] {\scalebox{2.5}{$X_1,\dots,X_4$}} (\ECCStart, -3);

        \draw[-{Stealth[scale=2]}, very thick] (\ECCStart-5, -6) -- node[above, anchor=south, xshift=-2mm] {\scalebox{2.5}{$X_1,\dots,X_4$}} (\ECCStart, -6);

        \draw[-{Stealth[scale=2]}, very thick] (\ECCStart-5, -9) -- node[above, anchor=south, xshift=-2mm] {\scalebox{2.5}{$X_1,\dots,X_4$}} (\ECCStart, -9);

        \draw[-{Stealth[scale=2]}, very thick] (\ECCStart-5, -12) -- node[above, anchor=south, xshift=-2mm] {\scalebox{2.5}{$X_1,\dots,X_4$}} (\ECCStart, -12);

        \node[anchor=north west, draw, minimum width=26cm, minimum height = 3cm, rotate=90] at (\ECCStart, -14) {\scalebox{4}{\texttt{Error Correction}}};
        \draw[-{Stealth[scale=2]}, very thick] (\ECCStart+3, -1) -- node[below, anchor=north, rotate=90, yshift=-1cm] {\scalebox{4}{\texttt{Codeword out}}} (\ECCStart+5, -1);             
        \draw[-{Stealth[scale=2]}, very thick] (-2, -13.5) -- node[below, anchor=north, xshift=-22cm] {\scalebox{4}{\texttt{Codeword in}}} (\ECCStart, -13.5);
        \draw[-{Stealth[scale=2]}, very thick] (-1, -13.5) -- (-1, 0) -- (0, 0);





\end{tikzpicture}

%% file: Sections/performance_analysis.tex
\section{Performance Analysis}
\label{results}

{
\setlength{\tabcolsep}{4pt}       
\begin{table}
    \centering
    \caption{Performance of our decoders synthesized on the XCZU48DR FPGA}
    \label{tab:fpga_results}
    \begin{tabular}{|c|c |c|c|c|c| c|c|c|c|}
    \hline
    \multirow{3}{*}{$n$} & \multirow{3}{*}{$t$} & \multicolumn{4}{c}{Conventional} & \multicolumn{4}{|c|}{Direct}\\
    \cline{3-10}
    & & $f_{\text{max}}$ & Lat. & \multirow{2}{*}{LUTs} & \multirow{2}{*}{Regs} & $f_{\text{max}}$ & Lat. & \multirow{2}{*}{LUTs} & \multirow{2}{*}{Regs} \\
    & & (MHz) & (ns) & & & (MHz) & (ns) & & \\
    \hline
    \multirow{3}{*}{$16$} & $2$ & 775 & 7.7 & 136 & 136 & 775 & 3.9 & 108 & 96\\
                          & $3$ & 775 & 10 & 284 & 267 & 775 & 5.2 & 271 & 163\\
                          & $4$ & 625 & 16 & 579 & 430 & 555 & 14 & 1000 & 294 \\
    \hline
    \multirow{3}{*}{$64$} & $2$ & 625 & 9.6 & 694 & 671 & 775 & 3.9 & 382 & 310 \\
                          & $3$ & 455 & 18 & 987 & 601 & 714 & 5.6 & 720 & 455 \\
                          & $4$ & 415 & 24 & 1524 & 865 & 500 & 16 & 2575 & 593\\
    \hline
    \multirow{3}{*}{$512$}& $2$ & 285 & 21 & 6164 & 3043 & 465 & 6.5 & 2397 & 2291 \\
                          & $3$ & 220 & 36 & 13864 & 3276 & 400 & 10 & 4596 & 2876 \\
                          & $4$ & 165 & 60 & 16065 & 8234 & 285 & 28 & 14436 & 2670 \\
    \hline
    
    \end{tabular}

\end{table}
}

When analyzing the performance of the proposed architecture, we consider three performance metrics: resource utilization, latency, and throughput. 
We synthesized our implementation for the Xilinx Ultrascale+ XCZU48DR FPGA and 16~nm FINFET. 
For resource utilization, we consider the number of LUTs and registers in FPGA, shown in Table~\ref{tab:fpga_results}. 
For ASIC synthesis, we consider the area in $\mu \text{m}^2$, shown in Fig.~\ref{fig:ASIC_area}. For ASIC, all architectures were synthesized $1$~GHz. 

The data in Table~\ref{tab:fpga_results} and Fig.~\ref{fig:ASIC_area} show that the direct architecture outperforms the conventional architecture in all metrics for $t=2$ and $t=3$, for all blocklengths. 
For $t=4$ the conventional architecture performs better for shorter blocklengths, while the direct architecture starts to perform better as the blocklength increases. The inflection point for the FPGA is around $n=64$, where the direct architecture still performs worse in terms of LUT utilization but better in terms of maximum frequency and registers. The post-synthesis ASIC results shows the inflection point at $n=256$. Regarding latency, the direct architecture performs better for all blocklengths analyzed.

BCH codes of length $256$ are often used in GPCs for optical communications~\cite{Wang2023JLT}. A more in-depth view for this group of BCH codes and a comparison with an implementation from the literature are shown in Table~\ref{tab:performance_256}. 
This table indicates that, for $n=256$, the direct decoder outperforms the conventional decoder in all respects for $t=2$ and $t=3$, and in most respects for $t=4$.
In FPGA, for $t=2$, we achieve $62\%$ lower latency, $35\%$ higher clock speed, $65\%$ lower LUT use, and $35\%$ lower register use. 
For $t=3$, we achieve $71.4\%$ lower latency, $75\%$ higher clock frequency, $54\%$ lower LUT use, and $45\%$ lower register use. 
For our post-synthesis ASIC evaluations, the direct decoder achieves $39\%$ and $35\%$ area savings while reducing latency by $50\%$ for $t=2$ and $t=3$, respectively.
For $t=4$, the performance gap closes between the conventional and direct decoder. The FPGA results show that our direct decoder has $35\%$ lower latency, $23\%$ higher clock frequency, $10\%$ lower LUT use and $58\%$ lower register use. For the post-synthesis ASIC results, the $t=4$ decoder achieves $25\%$ lower latency but needs $3\%$ more area. 

We compare our conventional decoder with the conventional architecture from~\cite{Ou2025CICC}, which uses no pipelining. Our decoder achieves significantly higher throughput at a slightly larger area. Post-synthesis results show we achieve an expected area efficiency of $10$~Tb/s/$\text{mm}^2$, compared to \cite{Ou2025CICC} at $1.9$~Tb/s/$\text{mm}^2$. Our conventional architecture is then expected to have $5$ times better area efficiency. It should be noted that the data from~\cite{Ou2025CICC} are from a physical chip, which has an increased overhead.

{
\setlength{\tabcolsep}{3.5pt}       
\begin{table}[]
    \centering
    \caption{Performance of BCH codes with blocklength $n=256$}
    \label{tab:performance_256}
    \begin{tabular}{|c|c|c |c|c|c|c |c|c|c|c|}
    \hline
    & \multirow{3}{*}{$t$} & & \multicolumn{4}{c}{XCZU48DR} & \multicolumn{4}{|c|}{16nm FinFET}\\
    \cline{4-11}
    &  & Lat.  & $f_{\text{max}}$ & Lat. &  \multirow{2}{*}{LUTs} & \multirow{2}{*}{Regs}  & $f_{\text{max}}$& Lat. & Area & Thr.\\
    &  & (cy.) & (MHz) & (ns) & & & MHz & (ns) & $\mu \text{m}^2$ & Gb/s \\
    \hline
    \multirow{4}{*}{\rotatebox[origin=c]{90}{Conv.}}& $2$ & 6 & 370 & 16 & 2926 & 1757 & 1000 & 6 & 4648 & 239\\
    & $3$ & 8 & 285 & 28 & 4999 & 2722 & 1000 & 8 & 8067 & 231\\
    & $4$ & 10 & 270 & 37 & 8463 & 3601 & 1000 & 10 &  12714 & 223\\
    & $6$ & 14 & 140 & 99 & 14912 & 2357 & 1000 & 14 & 20515 & \textbf{207} \\
    \hline
    \multirow{3}{*}{\rotatebox[origin=c]{90}{Direct}} & $2$ & 3 & 500 & 6 & 1018 & 1142 & 1000 & 3 & 2851 & 239 \\
    & $3$ & 4 & 500 & 8 & 2295 & 1498 & 1000 & 4 & 5277 & 231 \\
    & $4$ & 8 & 333 & 24 & 7631 & 1500 & 1000 & 8 &  13118 & 223 \\
    \hline
    \cite{Ou2025CICC} & $6$ & 15 & n/a\footnotemark[1] & n/a\footnotemark[1]  & n/a\footnotemark[1] & n/a\footnotemark[1] & 1040 & 14.4  & 16000 & \textbf{30.75}\\
    
    \hline
    \end{tabular}
\end{table}
}

\footnotetext[1]{No FPGA implementation results were reported in~\cite{Ou2025CICC}.}


%% file: Sections/conclusions.tex
\section{Conclusions}
\label{conclusions}
We implemented two decoder architectures for binary BCH codes. One based on the conventional BM algorithm in combination with Chien search and one based on directly computing the ELP and its roots. 
We provide an analysis of the performance of the two architectures, which shows that the direct decoder outperforms the conventional decoder for all metrics for $t=2$ and $t=3$. For $t=4$, the conventional decoder can still perform better for short blocklengths in terms of area. However, for longer blocklengths this advantage disappears and the direct decoder performs better for all metrics.
Lastly, our post-synthesis results for 16~nm FinFET show that our direct architecture achieves $239$ Gb/s throughput at $3$~ns latency for the $(256,239)$ eBCH code and $223$ Gb/s throughput at $8$~ns latency for the $(256, 223)$ eBCH code. 